\documentclass[a4paper,fleqn]{cas-dc}

\usepackage[numbers]{natbib}

\usepackage{xcolor}
\usepackage{subfig}
\usepackage{graphicx}
\usepackage{colortbl}
\usepackage{enumitem}
\usepackage{tabularx} 
\usepackage{geometry}

\usepackage{subcaption} 

\def\tsc#1{\csdef{#1}{\textsc{\lowercase{#1}}\xspace}}
\tsc{WGM}
\tsc{QE}

\begin{document}
\let\WriteBookmarks\relax
\def\floatpagepagefraction{1}
\def\textpagefraction{.001}

\shorttitle{}    

\shortauthors{}  

\title []{
RAG Safety: Exploring Knowledge Poisoning Attacks to Retrieval-Augmented Generation
}  



%

\author[1]{Tianzhe Zhao}[orcid=0000-0002-2879-2703]

\cormark[1]


\ead{ztz8758@foxmail.com}


\credit{Data curation,
Investigation,
Methodology,
Software,
Writing – original draft}

\affiliation[1]{organization={School of Computer Science and Technology,Xi'an Jiaotong University},
            city={Xi'an},
            postcode={710049}, 
            country={China}}

\author[2]{Jiaoyan Chen}


\ead{jiaoyan.chen@manchester.ac.uk}


\credit{Conceptualization,
Methodology,
Supervision,
Writing – review \& editing}

\affiliation[2]{organization={Department of Computer Science, The University of Manchester},
            city={Manchester},
            postcode={M13 9PL}, 
            country={United Kingdom}}

\author[1]{Yanchi Ru}
\ead{2196123580@stu.xjtu.edu.cn}
\credit{Data curation, Visualization}

\author[1]{Haiping Zhu}
\ead{zhuhaiping@xjtu.edu.cn}
\credit{Funding acquisition, Supervision}

\author[3]{Nan Hu}
\ead{nanhu@seu.edu.cn}
\credit{Investigation}
\affiliation[3]{organization={School of Computer Science and Engineering, Southeast University},
            city={Nanjing},
            postcode={211102}, 
            country={China}}

\author[1]{Jun Liu}
\ead{jiukeen@xjtu.edu.cn}
\credit{Funding acquisition, Supervision}

\author[4]{Qika Lin}[orcid=0000-0001-5650-0600]
\cormark[1]
\ead{qikalin@foxmail.com}
\credit{Investigation,
Methodology,
Writing – review \& editing}
\affiliation[4]{organization={National University of Singapore},
            city={Singapore},
            postcode={119077}, 
            country={Singapore}}

\cortext[1]{Corresponding author}



\begin{abstract}
Retrieval-Augmented Generation (RAG) enhances large language models (LLMs) by retrieving external data to mitigate hallucinations and outdated knowledge issues. Benefiting from the strong ability in facilitating diverse data sources and supporting faithful reasoning, knowledge graphs (KGs) have been increasingly adopted in RAG systems, giving rise to KG-based RAG (KG-RAG) methods. Though RAG systems are widely applied in various applications, recent studies have also revealed its vulnerabilities to data poisoning attacks, where malicious information injected into external knowledge sources can mislead the system into producing incorrect or harmful responses. However, these studies focus exclusively on RAG systems using unstructured textual data sources, leaving the security risks of KG-RAG largely unexplored, despite the fact that KGs present unique vulnerabilities due to their structured and editable nature.
In this work, we conduct the first systematic investigation of the security issue of KG-RAG methods through data poisoning attacks. To this end, we introduce a practical, stealthy attack setting that aligns with real-world implementation. We propose an attack strategy that first identifies adversarial target answers and then inserts perturbation triples to complete misleading inference chains in the KG, increasing the likelihood that KG-RAG methods retrieve and rely on these perturbations during generation. Through extensive experiments on two benchmarks and four recent KG-RAG methods, our attack strategy demonstrates strong effectiveness in degrading KG-RAG performance, even with minimal KG perturbations. In-depth analyses are also conducted to understand the safety threats within the internal stages of KG-RAG systems and to explore the robustness of LLMs against adversarial knowledge.
\end{abstract}





\maketitle

\section{Introduction}
\label{intro}
Retrieval-Augmented Generation (RAG) extends Large Language Models (LLMs) with access to external knowledge sources, enabling responses to be grounded in retrieved contents rather than generated solely from the model's internal parameters~\cite{gao2023retrieval,fan2024survey, lewis2020retrieval}. This approach helps mitigate several key limitations of LLMs, such as hallucinations, outdated knowledge, and weak domain adaptation. As a result, RAG has been widely adopted in a broad range of tasks, such as question answering~\cite{siriwardhana2023improving} and scientific summarization~\cite{toney-etal-2025-expertly}. Despite their success, traditional RAG systems still suffer from some issues like redundant information in the retrieved data and too long prompts, especially facing heterogeneous and large-scale data sources \cite{peng2024graph,liu-etal-2024-lost}.

Recently, knowledge graphs (KGs), which represent relational facts in the form of triples, i.e., \textit{(head, relation, tail)}, have been introduced for supporting RAG, giving rise to the direction of KG-based Retrieval-Augmented Generation (KG-RAG)~\cite{peng2024graph, han2024retrieval}. 
As a widely used approach for knowledge representation, there have been quite a few existing KGs (e.g., Wikidata \cite{vrandevcic2014wikidata} and SNOMED CT \cite{donnelly2006snomed}) that contain a large quantity of valuable knowledge. They can be directly used as effective sources for supporting both general purpose and domain specific application~\cite{luo2024reasoning,wu2024cotkr}.
KG and its associated semantic technologies also exhibit strong capabilities in data integration and knowledge management~\cite{ji2021survey}, and thus can be used to facilitate the consolidation of diverse external data sources, such as documents and tables, into more focused and semantically coherent representations for supporting RAG~\cite{wang2024knowledge,edge2024local}.
In addition, KGs can be used to support faithful reasoning, especially in domains that require high trustworthiness, such as legal judgment~\cite{ijcai2022p765} and medical diagnosis~\cite{wu2023medical}. Retrieved evidence from KGs, e.g., explicit relational paths, can provide a reliable and interpretable justification for RAG. 
Due to all these benefits, KG-RAG has become a popular research topic with quite a few emerging methods.\footnote{Note that many KG-based RAG methods, which KGs as retrieval sources, are simply referred to as \textit{GraphRAG}~\cite{peng2024graph,han2024retrieval}, although the graphs they use fall into the scope of KG.}

As RAG systems become increasingly prominent in real-world applications, several studies~\cite{zou2024poisonedrag, nazary2025poison, zhou2025trustrag, tan-etal-2024-glue} have also witnessed their vulnerabilities, particularly when the external knowledge sources are of low quality or have been adversarially manipulated.
They find that injecting malicious or misleading content into the external knowledge source can cause RAG systems to generate incorrect or even specified responses, without modifying the LLM parameters. These manipulations, commonly referred to as \textit{data poisoning attacks}~\cite{fan2022survey, wang2022threats}, introduce adversarial perturbations into the data source. However, existing studies have only focused on the poisoning of traditional RAG systems that leverage unstructured text such as Wikipedia pages~\cite{zou2024poisonedrag} or web documents~\cite{tan-etal-2024-glue}, leaving the security risks of KG-RAG systems largely unexplored.

\begin{figure*}[h]
  \centering
  \includegraphics[width=0.95\linewidth]{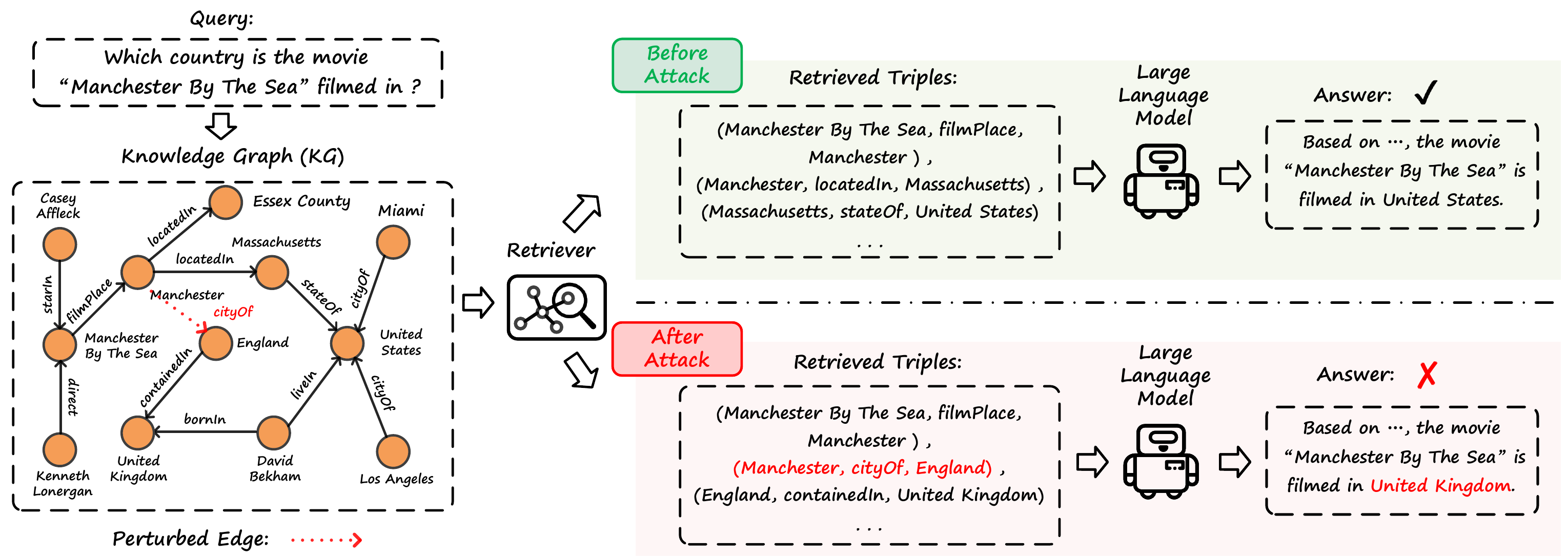}
  \caption{Illustration of KG-RAG pipelines before and after data poisoning attack.}
  \label{fig:intro}
\end{figure*}

Although the integration of KGs can enable faithful reasoning through explicit inference chains~\cite{luo2024reasoning, luo2024graph}, KG-RAG systems are also exposed to unique vulnerabilities. 
In practice, many widely-used KGs, such as Wikidata~\cite{vrandevcic2014wikidata}, collect data from public resources and can be edited by users. This openness significantly lowers the barrier for attackers to insert poisoned triples, making stealthy poisoning feasible at low cost~\cite{you2023mass, 10.1145/3626772.3657702}. 
Even in internally-curated KGs, data perturbations can still be hardly avoided during automatic KG construction, such as noisy information extraction~\cite{bratus2011domain} and imperfect entity linking~\cite{shen2014entity}.
Moreover, the inherent structure of KGs makes them highly sensitive to small perturbations. Even minor modifications, such as adding several perturbation triples, can significantly alter both the local and global semantics of the graph~\cite{ijcai2019p674, bhardwaj-etal-2021-poisoning, you2023mass, 10.1145/3626772.3657702}. These structural shifts may mislead the retriever into selecting incorrect triples, ultimately leading to an incorrect response. As illustrated in Figure~\ref{fig:intro}, inserting a single edge between \textit{Manchester} and \textit{England} can cause KG-RAG to incorrectly answer \textit{United Kingdom} to the question “Which country is the movie \textit{Manchester By The Sea} filmed in?”, by activating a misleading inference chain. These observations highlight the susceptibility of KG-RAG systems to data poisoning attacks.
Therefore, investigating how KG-RAG methods are affected when the KGs to retrieve are polluted is both timely and urgent for understanding their robustness and security threats.

To fill this gap, we conduct the first systematic study of data poisoning attacks on KG-RAG methods. Specifically, we focus on attacks that inject a small number of adversarial triples into the underlying KG to manipulate multi-hop reasoning and induce incorrect outputs.
To ensure that the attack remains both realistic and stealthy, we build a constrained threat model aligned with real-world KG-RAG deployment scenarios. 
The attack operates under a black-box setting, in which the attacker has no access to the retriever, the language model, or any internal parameters of the KG-RAG method. We further assume that the attacker only knows the question for attacking, but without the knowledge of its ground-truth answers.
Regarding KG modification, the attacker is restricted to inserting a small number of triples, as many real-world KGs (e.g., Wikidata) require additional permissions for editing or deletion. Moreover, insertion-based attack offers better stealth and feasibility in practice.
To further improve stealth, we assume that all the inserted triples are composed of entities and relations already present in the KG, avoiding the introduction of new entities or relations that are more likely to be detected.

Based on the above settings, we propose an effective data poisoning attack strategy, with the aim of constructing misleading inference chains that guide KG-RAG methods toward incorrect answers. Given a user question, we first generate several plausible but incorrect answers using a general-purpose LLM. These incorrect answers are aligned to existing KG entities to avoid introducing new entities, yielding a set of targeted incorrect entities that serve as attacker-intended responses to the question. To facilitate the creation of misleading inference chains, a KG-specific language model~\cite{luo2024reasoning} is employed to generate relation paths that are structurally or semantically associated with the question, which act as the templates for the instantiation of inference chains in the KG. The final perturbation triples are constructed to complete the misleading chains that originate from the question topic entity and terminate at the targeted incorrect entities.

We conduct comprehensive experiments to evaluate the effectiveness of our attack strategy on two knowledge graph-based question answering (KGQA) benchmarks against four recent KG-RAG methods. Experimental results show that even minimal insertion of perturbation triples can significantly degrade the reasoning performance of KG-RAG methods, highlighting the vulnerability of existing KG-RAG methods to data poisoning attacks. 
To obtain a better understanding of the safety issue, we conduct in-depth analyses of how KG poisoning impacts both the retrieval and generation stages.
Our findings indicate that the primary vulnerability lies in the retrieval stage, where injected perturbation triples achieve remarkably high coverage, with at least one adversarial triple being retrieved in over 90\% of the questions under attack.  
However, the inclusion of perturbations in the retrieval results does not necessarily lead to their incorporation into the generated answers, as LLMs may selectively filter out adversarial content based on their internal knowledge. 
Consequently, we further examine the robustness of different LLMs to poisoned retrieval results and identify an interesting trend: LLMs with stronger KG reasoning capabilities are also more sensitive to KG poisoning.

The main contributions of this work can be summarized as:

\noindent $\bullet$ We investigate the safety risks of KG-RAG systems, and present the first systematic study of data poisoning attacks under realistic, black-box settings.

\noindent $\bullet$ We develop a targeted attack strategy that injects perturbation triples into the KG to complete adversarial inference chains, thereby misleading KG-RAG methods into producing incorrect answers.

\noindent $\bullet$ We conduct extensive experiments on two benchmarks and four representative KG-RAG methods. The results demonstrate that our attack substantially degrades the performance of existing systems. In-depth analyses provide a comprehensive understanding of the security vulnerabilities of KG-RAG systems and offer actionable insights for enhancing their robustness.

\section{Background}
\subsection{Preliminaries}
\label{sec:prelim}
\noindent \textbf{Knowledge Graphs (KGs)} represent factual knowledge in a structured format, typically defined as $\mathcal{G} = (\mathcal{E}, \mathcal{R}, \mathcal{T})$, where $\mathcal{E}$ is a set of entities, $\mathcal{R}$ a set of relations, and $\mathcal{T} \subseteq \mathcal{E} \times \mathcal{R} \times \mathcal{E}$ is a set of factual triples. Each triple $(e_h, r, e_t)$ in $\mathcal{T}$ denotes a relation $r$ linking a head entity $e_h$ to a tail entity $e_t$.

\noindent \textbf{Reasoning Paths} are widely-used to support multi-hop reasoning over KGs. A reasoning path is typically formed as a sequence of consecutive triples in the KG, denoted as:
\[
    p: e_0 \xrightarrow{r_1} e_1 \xrightarrow{r_2} \dots \xrightarrow{r_l} e_l
\]
where $(e_{i-1}, r_i, e_i) \in \mathcal{T}$, and $l$ is the length.
Such paths capture compositional semantics and serve as interpretable evidence for multi-hop inference in KG-based applications. 
For example, given the reasoning path: \textit{Cardiff} $\xrightarrow{locatedIn}$ \textit{Wales} $\xrightarrow{containedIn}$ \textit{United Kingdom}, we could conclude that \textit{Cardiff} is a city in the \textit{United Kingdom}.

\noindent \textbf{Relation Paths} abstract away the intermediate entities in a reasoning path while preserving its relational structure. For each reasoning path $p$, we denote its associated relation path as $w = (r_1, r_2, \dots, r_l)$. For instance, \textit{(locatedIn, containedIn)} is the corresponding relation path of the aforementioned reasoning path from \textit{Cardiff} to \textit{United Kingdom}.
Relation paths are widely used for deriving compositional patterns across KGs~\cite{sun2018rotate, cheng2023neural}. 
Conversely, given a relation path $w$, we can infer matching reasoning paths from a KG that instantiate $w$ by identifying valid triple sequences. We refer to this procedure as \textit{grounding} in this work.

\noindent \textbf{Knowledge Graph-based Question Answering (KGQA)} serves as a cornerstone downstream task for KG-RAG~\cite{peng2024graph}, and is predominantly used to evaluate the KG-RAG's ability to retrieve and reason over KGs. Given a natural language question $q$ and a KG $\mathcal{G}$, the task aims to design a function $f$ that utilizes $\mathcal{G}$ to return an answer entity $a \in \mathcal{E}$, i.e., $a = f(q, \mathcal{G})$.




\subsection{Knowledge Graph-based Retrieval-Augmented Generation}
\subsubsection{General Framework}
Knowledge graph-based retrieval-augmented generation (KG-RAG) is a framework that enhances LLMs by leveraging external KGs to support factual, interpretable, and multi-hop reasoning. 
Given a user question \( q \) and a KG \( \mathcal{G} \), KG-RAG systems typically follow a two-stage pipeline:

\textbf{Retrieval stage} focuses on identifying a sub-KG \( \mathcal{G}_q = (\mathcal{E}_q, \mathcal{R}_q, \mathcal{T}_q)\) from  \(\mathcal{G} \) that is most relevant to the input question:
\begin{equation}
    \mathcal{G}_q = \text{Retriever}(q; \mathcal{G}).
\end{equation}
Various strategies have been proposed to implement the retriever in existing KG-RAG methods. Some approaches retrieve question-relevant triples based on embedding similarity~\cite{wu-etal-2024-cotkr}, though the retrieved triples may not form a connected subgraph. To improve structural coherence, heuristic rules~\cite{NEURIPS2024_efaf1c97} are applied to further select connected triples and construct meaningful subgraphs. Another popular design leverages the strong generative capabilities of LLMs to guide the retrieval in reasoning paths~\cite{luo2024reasoning}.

\textbf{Generation stage} involves synthesizing responses based on the retrieved graph data: $a = f_\theta(q, \mathcal{G}_q),$
where \( f_\theta \) denotes the LLM generator parameterized by \( \theta \).  
In practice, a common approach is to prompt the LLM with a serialized natural language input that combines the question \( q \) and the retrieved content \( \mathcal{G}_q \). Thus, this stage can be denoted as:
\begin{equation}
    a = f_\theta(\text{Prompt}[q, \mathcal{G}_q]).
\end{equation}




\subsubsection{KG-RAG Methods}
\label{sec: kg-rag_methods}
In this work, we explore data poisoning attacks on four representative KG-RAG methods: RoG~\cite{luo2024reasoning}, RCR~\cite{luo2024graph}, G-retriever~\cite{NEURIPS2024_efaf1c97}, and SubgraphRAG~\cite{li2025simple}. These methods primarily focus on retrieving high-quality relevant triples to support accurate answer generation. Based on their retrieval structures, the four KG-RAG methods can be broadly categorized into two groups: path-retrieval methods (RoG and GCR) and subgraph-retrieval methods (G-retriever and SubgraphRAG). More details are introduced as follows:

\textbf{RoG} extends the basic KG-RAG framework with a two-module design: \textit{planning} and \textit{retrieval-reasoning}. Given a question, RoG first employs a simple instruction template to prompt LLMs to generate relation paths that can be grounded in the KG as faithful plans through the \textit{planning} module. Subsequently, the \textit{retrieval-reasoning} module grounds these relation paths to retrieve corresponding reasoning paths from the KG and performs reasoning via LLM prompting. Since general LLMs lack prior knowledge of the KG’s relational facts, directly using them may lead to unfaithful relation path generation and incorrect reasoning. To address this, RoG applies an instruction-tuning strategy to integrate KG knowledge into the LLM, developing a KG-specific LLM \footnote{This enhanced LLM model is also denoted as RoG in the original work. To avoid confusion, we denote the model as LLM$_\text{RoG}$ in this paper.} used for both relation path generation and reasoning.

\textbf{GCR} enhances the KG-RAG framework by introducing a graph-constrained decoding strategy. It adopts a breadth-first search (BFS) algorithm to retrieve reasoning paths within a predefined length starting from the entity mentioned in the question. These paths are transformed into token sequences and stored in a specialized prefix tree structure called KG-Trie. 
GCR then utilizes a lightweight, KG-specific LLM together with the KG-Trie to generate multiple reasoning paths along with hypothesis answers that the paths can infer. 
The KG-Trie guides and restricts the LLM, ensuring it produces only valid reasoning paths that can be grounded in the KG. 
These generated reasoning paths and their hypothesis answers are then passed to a general-purpose LLM for producing accurate and reliable final answers.

\textbf{G-retriever} performs KG-augmented retrieval by encoding the question, as well as the entities and relations in the KG, using SentenceBERT~\cite{reimers2019sentence}. It retrieves the top-k most relevant entities and relations via searching k nearest neighbors based on the cosine similarity. 
A connected subgraph is constructed by heuristically selecting top-ranked nodes and edges, which are assigned relevance scores based on similarity values, while controlling the graph size via edge cost constraints.
The resulting subgraph is encoded using a Graph Attention Network (GAT), and the output embedding is used as a soft prompt for a decoder-only LLM to generate the final answers.

\textbf{SubgraphRAG} first formulates subgraph retrieval as a distribution optimization problem and adopts a retriever that decomposes subgraph construction into independent selections over triples, allowing each triple to be scored and optimized separately to improve retrieval efficiency.
To capture the structural relevance between the question and each triple, it computes directional distance encodings that reflect the position of the triple relative to the topic entities in the question. These structural features are concatenated with the question representation and the semantic representation of the triple, and then passed through a lightweight multilayer perceptron to estimate the inclusion probability of each triple. The top-ranked triples are selected to form a subgraph of adaptive size, which is then linearized into a text sequence and incorporated into the LLM input prompt for in-context learning.

\subsection{Threat Model}
\label{sec: threat_model}
This section formalizes the threat model under which our data poisoning attack operates, including the attacker's objectives, knowledge assumptions, and capabilities.

\subsubsection{Attacker's Objective}
The attacker aims to subvert the behavior of KG-RAG methods by injecting perturbation triples into the external KG $\mathcal{G}$. 
Formally, given a user question $q$, the attacker seeks to insert a small set of adversarial triples $\hat{\mathcal{T}}$ into $\mathcal{G}$ such that the KG-RAG method generates a set of incorrect answers $\hat{A}$ that deviates from the correct gold answer $a^*$:
\begin{equation}
    \hat{A} = \text{KG-RAG}(q; \hat{\mathcal{G}}), \quad a^* \notin \hat{A},
\end{equation}
where $\hat{\mathcal{G}} = \{ \mathcal{E}, \mathcal{R}, \mathcal{T} \cup \hat{\mathcal{T}}\}$.

\subsubsection{Attacker's Knowledge}
We assume a black-box setting where the attacker has the following knowledge:

\noindent $\bullet$ \textbf{Access to the External Data Source:}
The attacker can inspect and modify the KG $\mathcal{G}$ used by the KG-RAG method, including its entities, relations, and topological structure. 

\noindent $\bullet$ \textbf{No Access to KG-RAG Internal Modules:}
The attacker has no access to the internal modules of the KG-RAG method, including the retriever, the language model (both its architecture and parameters), or the prompt templates. Furthermore, the attacker is unaware of the ground-truth answers to any questions. All attack actions are executed without observing KG-RAG outputs or internal signals, making the attack scenario fully black-box.

This setting reflects a practical and realistic attack surface, where only the external KG is visible and editable. 
Following prior work on attacks against text-based RAG~\cite{zou2024poisonedrag, tan-etal-2024-glue}, we note that the LLM employed internally by the target KG-RAG method can still be used to build the attacker.
%
This does not violate the aforementioned black-box assumption, as the attacker does not interact with or observe the internal behavior of the deployed KG-RAG system.

\subsubsection{Attacker's Capabilities and Constraints}
In this study, the attacker is limited to \textbf{insertion-only} perturbations, i.e., they can only inject a small number of perturbation triples $\hat{\mathcal{T}}$ into the external KG, but cannot delete or modify existing facts. 
This setting makes the attack more stealth and feasible in many real-world KGs.

Furthermore, considering the stealthiness and practicality required in real-world scenarios, the attack is further constrained by the following conditions:

\noindent $\bullet$ All inserted triples must be composed of existing entities and relations in the KG $\mathcal{G}$ to mitigate the risk of being detected. 
Under this constraint, each perturbation triple is defined as $(\hat{e}_h, \hat{r}, \hat{e}_t) \notin \mathcal{T}$, where $\hat{e}_h, \hat{e}_t \in \mathcal{E}$ and $\hat{r} \in \mathcal{R}$.

\noindent $\bullet$ The number of injected triples for each question is limited to at most $K$ per question, where $K \ll \lvert \mathcal{T} \rvert$, reflecting realistic attack budgets and constraining the perturbation scale.

\begin{figure*}[h]
  \centering
  \includegraphics[width=0.95\linewidth]{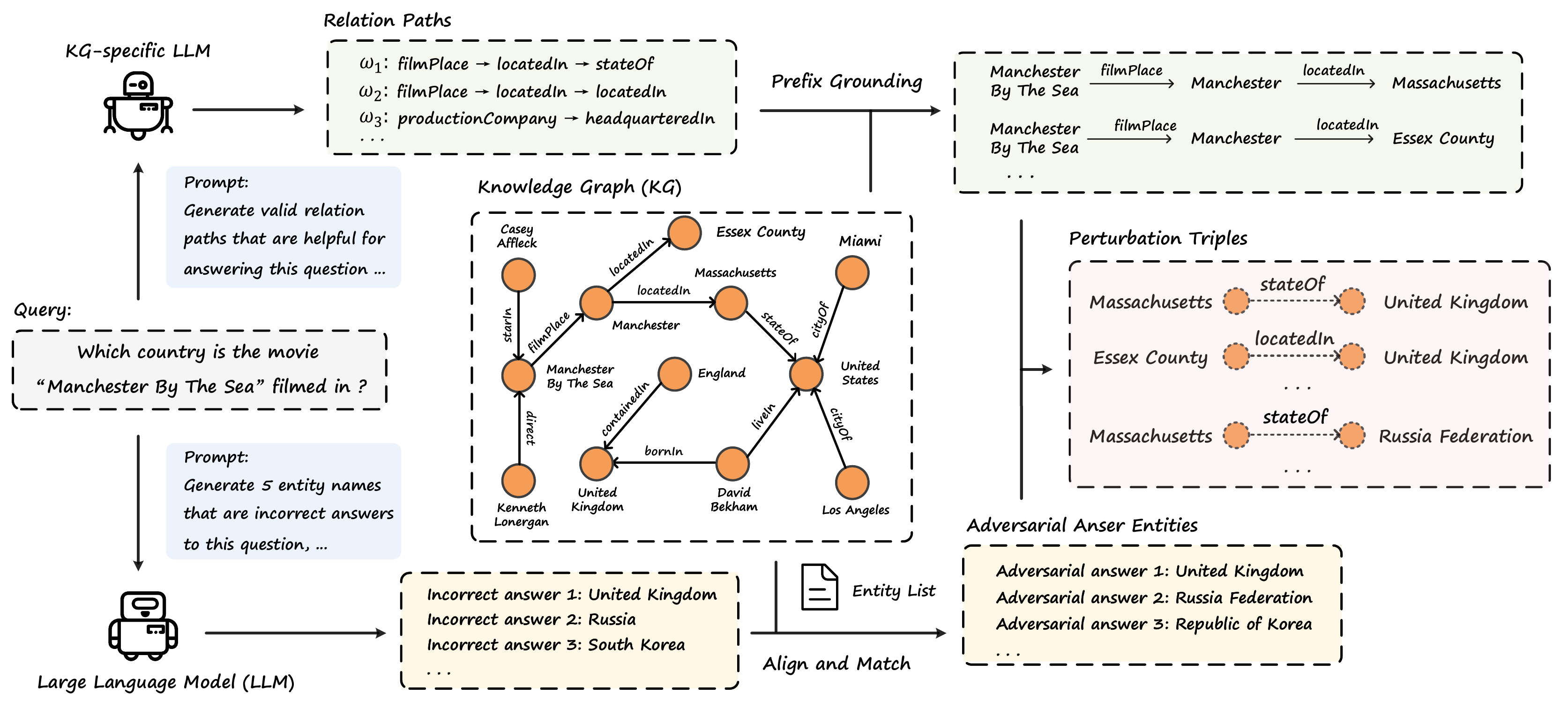}
  \caption{The overall framework of our proposed data poisoning attack against KG-RAG.}
  \label{fig:method}
\end{figure*} 

\section{Attack Method}
In this section, we detail our data poisoning attack strategy, which inserts perturbation triples into the KG to 
steer KG-RAG methods toward generating incorrect answers. The overall framework is illustrated in Figure~\ref{fig:method}.

\subsection{Adversarial Answer Generation}
A key challenge in attacking a KG-RAG method lies in distinguishing attack-induced failures from errors caused by the features of the method itself, such as its uncertainty and LLM hallucination. 
To address this, we assign each question a set of adversarial answers, which serve as explicit attack targets. Our attack strategy aims to guide KG-RAG methods to generate these answers in place of the correct ones, allowing for the clear determination of whether the attack has succeeded.

Given a user question $q$, to improve the chances that the associated perturbation triples are retrieved, but also enhance the likelihood that the corresponding adversarial answers ${\hat{A}}$ of these perturbation triples are favored by the LLM during generation, ${\hat{A}}$ are expected to align with the question’s implicit semantic constraints. For instance, when attacking the question ``\textit{Which country is Manchester By The Sea filmed in}'',  a suitable adversarial answer should correspond to a country or location. In contrast, if the adversarial answer refers to a semantically irrelevant type, such as a person or an event, it is likely to be excluded by the LLM during generation based on its internal knowledge, even if the associated triples are successfully retrieved. Similar semantic constraints are adopted in poisoning attacks on KG completion~\cite{you2023mass}, where entity type information from rich ontology knowledge is used to ensure that the disrupted entities remain type-consistent with the original targets. 
However, in real-world scenarios, it is not practical to acknowledge the ground-truth answers in advance and further enforce such type consistency through explicit supervision.

\begin{figure}
    \centering
    \includegraphics[width=\linewidth]{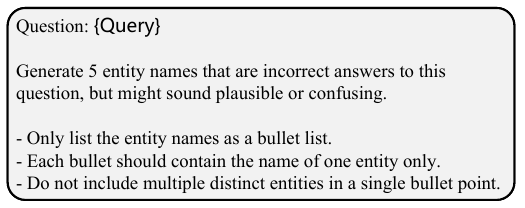}
    \caption{Prompt template for generating adversarial answers.}
    \label{fig:prompt1}
\end{figure}

To this end, we design a prompting strategy that guides an LLM to produce incorrect but semantically plausible answers, as illustrated in Figure~\ref{fig:prompt1}. We adopt a general-purpose LLM to ensure broad applicability across different KG-RAG methods.
Considering that the LLM may produce entities not present in the KG, we perform multiple rounds of generation for each question $q_i$ and apply fuzzy string matching to the KG entity list, ultimately determining $N$ final adversarial answers, denoted as $ \hat{\mathcal{A}}_i = \{ \hat{a}_{i,1}, \hat{a}_{i,2}, \ldots, \hat{a}_{i,N} \} $, where each $\hat{a}_{i,j} \in \mathcal{E}$ corresponds to valid entities in the KG $\mathcal{G}$.

\subsection{Relation Path Extraction}
\label{sec:rel_path}
To construct triples for effective poisoning, we extract relation paths that reflect plausible reasoning patterns that KG-RAG methods may follow when answering a given question. These relation paths are later utilized to construct adversarial reasoning paths that redirect the inference process toward the adversarial answers.

Despite the strong generalization and reasoning capabilities of general-purpose LLMs such as GPT-4, it is not an optimal solution to directly prompt them to derive relation paths, as they often lack explicit awareness of structured knowledge graphs and struggle to generate interpretable, KG-consistent relation paths. A more promising solution is to develop a KG-specific LLM that incorporates structural knowledge of the KG and is capable of generating plausible multi-hop relation paths.

This capability can be formalized as an optimization objective. Given a question $q$, the goal is to generate multiple relation paths that are both grounded in the KG and semantically aligned with answering $q$. To enable this, weak supervision is introduced via the shortest relation paths between the topic and answer entities in the KG, which reflect plausible reasoning trajectories over the KG structure, with the following optimization goal:
\begin{equation}
   \max_\theta \ \mathbb{E}_{w \sim \mathcal{W}^*} [log P_{\theta}(w \mid q)],
\end{equation}
where $\mathcal{W}^*$ denotes the set of shortest relation paths between the topic entity in the question and the gold answer entity in the KG, and $P_\theta(w \mid q)$ is the probability of generating a relevant relation path grounded by the KG. With the flexible assumption of a uniform distribution over $\mathcal{W}^*$, this expectation be approximated by:
\begin{equation}
   \arg\max_\theta \frac{1}{|\mathcal{W}^*|} \sum_{w \in \mathcal{W}^*} \log P_\theta(w \mid q).
   \label{eq: opt}
\end{equation}
Since the relation path is represented as a sequence of relations (as introduced in Section~\ref{sec:prelim}), the objective in Equation~\ref{eq: opt} can be further decomposed as:
\begin{equation}
    \frac{1}{|\mathcal{W}^*|} \sum_{w \in \mathcal{W}^*} \log \prod_{i=1}^{|w|} P_{\theta}(r_i \mid r_{<i}, q),
\end{equation}
This optimization formulation follows the core training principle of LLM$_\text{RoG}$~\cite{luo2024reasoning}, a publicly available KG-specific LLM trained with a combined objective of generating faithful multi-hop reasoning paths and producing final answer predictions. Under this formulation, a novel KG-specific LLM can also be trained from scratch to generate relation paths that serve as structural templates for perturbation insertion, guiding where and how adversarial triples should be introduced to misdirect the reasoning process.

\subsection{Perturbation Insertion}
For each question $q$, after obtaining its adversarial answers $\hat{\mathcal{A}}$ and the corresponding faithful and relevant relation paths $\mathcal{W}$, we construct perturbation triples that inject misleading reasoning chains into the KG.

Specifically, for each relation path $w = (r_1, r_2, \cdots, r_l)$, we first attempt to ground its prefix $w' = (r_1, r_2, \cdots, r_{l-1})$ in the KG, starting from the topic entity $e_q$:
\begin{equation}
    \mathcal{E}_{l-1} = \text{Grounding}(e_q, w'; \mathcal{G}),
\end{equation}
where $\text{Grounding}(\cdot)$ denotes the grounding operation, and $e_q$ is the topic entity in the question $q$. Each $e_{l-1} \in \mathcal{E}_{l-1}$ corresponds to a valid entity reached by the path:
\[
    e_q \xrightarrow{r_1} e_1 \xrightarrow{r_2} \dots \xrightarrow{r_{l-1}} e_{l-1}.
\]
We then construct perturbation triples of the form $(e_{l-1},\ r_l,\ \hat{a})$ by attaching each adversarial answer $\hat{a} \in \hat{\mathcal{A}}$.
To enhance the stealthiness of the attack, we limit the number of perturbation triples associated with each adversarial answer $\hat{a} \in \hat{\mathcal{A}}$ to a fixed budget $K$.

In practice, however, due to the possible sparse KG structure, some relation paths in $\mathcal{W}$ may either fail to ground any valid $(l-1)$-hop entity, or yield few triples to meet the perturbation budget $K$ for the adversarial answer. 
In such cases, we adopt a fallback strategy to generate additional perturbation triples that preserve the intended relation pattern by randomly sampling intermediate entities from the $\mathcal{G}$ to serve as connective bridges between $e_q$ and $\hat{a}$ along the given relation path. Taking a 2-hop relation path $(r_1, r_2)$ as example, two perturbation triples $(e_q, r_1, e')$ and $(e', r_2, \hat{a}_{i,j})$ are together inserted into the graph, where $e'\in \mathcal{E}$ is a bridge entity randomly sampled from the graph. 
This fallback strategy supplements the perturbation triples for each adversarial answer $\hat{a} \in \hat{\mathcal{A}}$ to ensure the number reaches $K$.

\section{Evaluation}

\subsection{Datasets}
We evaluate the effectiveness of our attack using two widely adopted KGQA benchmarks: WebQuestionSP (WebQSP)~\cite{yih2016value} and Complex WebQuestions (CWQ)~\cite{talmor-berant-2018-web}. Both datasets are built on Freebase and widely used to evaluate KG-RAG systems which usually first retrieve evidence from Freebase and then generate answers of the target natural language questions.
To ensure comprehensive and systematic evaluation, we use the entire test sets of WebQSP and CWQ as our attack targets. Following previous research on KG-RAG~\cite{luo2024reasoning, he2021improving}, we utilize the publicly available preprocessed versions of WebQSP\footnote{https://huggingface.co/datasets/rmanluo/RoG-webqsp} and CWQ\footnote{https://huggingface.co/datasets/rmanluo/RoG-cwq},  where each question-answer pair is associated with a subgraph of Freebase, constructed by extracting all triples within a specific distance from the top entity of the question. In this work, a data poisoning attack is applied directly on these preprocessed subgraphs. The statistics for these two datasets are presented in Table~\ref{tab:dataset1} and \ref{tab:dataset2}.

\begin{table}[]
\caption{Statistics of the WebQSP and CWQ test sets. \#Test denotes the number of questions in the set test; \#1-hop \slash 2-hop \slash $\ge$3-hop indicate the proportion of questions requiring reasoning over 1,2, or at least 3 hops, respectively; $|\bar{\mathcal{T}}|$ represents the mean size of subgraphs (in number of triples) associated with each question in the test set.}
\fontfamily{ptm}\selectfont
\resizebox{0.48\textwidth}{!}{
\begin{tabular}{cccccc}
\toprule
\textbf{Dataset} & \textbf{\#Test} & \textbf{\#1-hop} & \textbf{\#2-hop} & \textbf{\# $\geq$3-hop} & \textbf{$\bar{\lvert \mathcal{T}\rvert}$} \\
\midrule
WebQSP & 1,628 & 65.49\% & 34.51\% & 0 & 4,228 \\
CWQ    & 3,531 & 40.91\% & 38.64\% & 20.75\% & 4,195 \\
\bottomrule
\end{tabular}
}
\label{tab:dataset1}
\end{table}

\begin{table}[]
\caption{Statistics of the number of gold answers for the question in WebQSP and CWQ test sets.}
\fontfamily{ptm}\selectfont
\resizebox{0.48\textwidth}{!}{
\begin{tabular}{ccccc}
\toprule
       & \textbf{\#Ans=1} & \textbf{ 2 $\leq$ \#Ans $\geq$ 4} & \textbf{ 5 $\leq$ \#Ans $\geq$ 9} & \textbf{ \#Ans $\geq$ 10} \\
\midrule
WebQSP & 51.2\% & 27.4\% & 8.3\% & 12.1\% \\
CWQ    & 70.6\% & 19.4\% & 6.0\% & 4.0\% \\
\bottomrule
\end{tabular}
}
\label{tab:dataset2}
\end{table}

\subsection{Evaluation Metrics}
We evaluate the attacking and the KG-RAG methods from two perspectives:

\noindent \textbf{QA Performance Degradation.}
This perspective assesses the extent to which data poisoning undermines the QA capabilities of KG-RAG methods. We report standard evaluation metrics on both WebQSP and CWQ datasets, including Hit, Precision, Recall, F1, Hits@1, and Exact Match (EM).  Lower metric values under attack reflect a stronger degradation in QA performance, indicating higher attack effectiveness. Below, we briefly describe each metric:

\noindent $\bullet$ \textbf{Hit} is the proportion of the questions for which at least one gold answer appears in the predicted answer set, providing a coarse measure of KGQA performance.

\noindent $\bullet$ \textbf{Precision} measures the proportion of correct answers among all predicted answers, computed per question and averaged over the test set.

\noindent $\bullet$ \textbf{Recall} quantifies answer completeness by measuring the proportion of gold answers that are successfully generated for each question, with the same per-question averaging.

\noindent $\bullet$ \textbf{F1} is the harmonic mean of the above metrics, Precision and Recall, serving as an overall assessment towards the answer quality. 

\noindent $\bullet$ \textbf{Hits@1} is the proportion of questions of which the highest-ranked predicted answer is among the gold answer set, emphasizing the method's ability to prioritize correct answers.

\noindent $\bullet$ \textbf{Exact Match (EM)} evaluates strict correctness by measuring the proportion of questions for which the predicted answer set exactly matches the gold answer set.

\noindent \textbf{Success of Adversarial Manipulation.} 
In addition to evaluating how KGQA results decrease under poisoning, we provide a complementary perspective that explicitly measures the tendency of KG-RAG methods to generate assigned adversarial answers. While Attack Success Rate (ASR) is frequently used to measure the fraction of questions whose predicted answers are adversarial targets, it fails to provide a comprehensive view of the answer distribution and model preference, especially in scenarios where multiple answers are predicted. To this end, we propose three novel metrics as follows:

\noindent $\bullet$ \textbf{Attack-oriented Precision (A-Precision)} quantifies how dominantly the adversarial answers are generated by KG-RAG methods under attack. Specifically, it is computed as the average ratio of the adversarial answers that are within the predicted answer set across all questions:
\begin{equation}
    \text{A-Precision} = \frac{1}{\lvert \mathcal{Q} \rvert} \sum_{i=1}^{\lvert \mathcal{Q} \rvert} \frac{|\mathcal{P}_i \cap \hat{\mathcal{A}}_i|}{|\mathcal{P}_i|},
\end{equation}
where $\lvert \mathcal{Q} \rvert$ denotes the number of questions in the test set, $\mathcal{P}_i$ is the predicted answer set for the $i$-th question, and $\hat{\mathcal{A}}_i$ denotes the adversarial target answers assigned to the $i$-th question during our attack strategy.

\noindent $\bullet$ \textbf{Attack-oriented Hits@1 (A-H@1)} measures the proportion of questions for which the highest-ranked predicted answer by the KG-RAG method under attack is an adversarial answer, which is defined as:
\begin{equation}
\text{A-H@1} = \frac{1}{\lvert \mathcal{Q} \rvert} \sum_{i=1}^{\lvert \mathcal{Q} \rvert} \mathbb{I}\left[\text{Top}_1(\mathcal{P}_i) \in \hat{\mathcal{A}}_i\right],
\end{equation}
where $\text{Top}_1(\mathcal{P}_i)$ denotes the top-1 predicted answer for the $i$-th question, and $\mathbb{I}[\cdot]$ is the indicator function.

\noindent $\bullet$ \textbf{Attack-oriented Mean Reciprocal Rank (A-MRR)} is designed to capture the average ranking position of adversarial answers in the KG-RAG's predictions when suffering a data poisoning attack. It is computed as the average reciprocal rank of the highest-ranked adversarial answer for each question over the test set:
\begin{equation}
    \text{A-MRR} = \frac{1}{\lvert \mathcal{Q} \rvert} \sum_{i=1}^{\lvert \mathcal{Q} \rvert} \frac{1}{\text{rank}(\hat{\mathcal{A}}_i, \mathcal{P}_i)},
\end{equation}
where $\text{rank}(\hat{\mathcal{A}}_i, \mathcal{P}_i)$ denotes the position of the highest-ranked adversarial answer from the target set $\hat{\mathcal{A}}_i$ within $\mathcal{P}_i$. If none of the adversarial answers appear in $\mathcal{P}_i$, the reciprocal rank is defined as 0.

These three metrics collectively measure the effectiveness of adversarial manipulation, with higher values indicating greater success in forcing KG-RAG methods to generate the targeted adversarial answers.

\begin{figure}
    \centering
    \includegraphics[width=\linewidth]{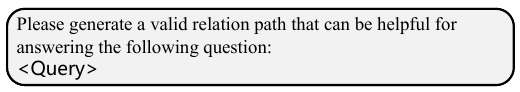}
    \caption{Prompt template for relation paths.}
    \label{fig:prompt2}
\end{figure}

\begin{table*}[t]
\fontfamily{ptm}\selectfont
\caption{KGQA performance ($\%$) under data poisoning attack. Lower values indicate better attack results. \textbf{Clean} denotes the original scenario without attack. \textbf{Bold} numbers denote optimal results. The \textcolor{red}{red} values in the subscripts denote the relative drops (compared to the Clean scenario) on the F1, Hits@1 and Exact Match(EM) metric. \emph{Due to the excessive data processing time required by G-retriever on CWQ (over 100 hours), we did not include its results on this dataset in our evaluation.}}
\resizebox{0.98\textwidth}{!}{
\fontsize{12}{16}\selectfont
\begin{tabular}{cccccccc|cccccc}
\toprule
\multirow{2}{*}{\textbf{KG-RAG}}      & \multirow{2}{*}{\textbf{Attacker}} & \multicolumn{6}{c|}{\textbf{WebQSP}}                                                        & \multicolumn{6}{c}{\textbf{CWQ}}                                                            \\ 
\cmidrule{3-14} 
                                      &                   & \textbf{Hit}  & \textbf{F1}  & \textbf{Precision} & \textbf{Recall} & \textbf{Hits@1} & \textbf{EM}                                                     &  \textbf{Hit} & \textbf{F1}  & \textbf{Precision} & \textbf{Recall} & \textbf{Hits@1} & \textbf{EM} \\ 
\midrule
\multirow{3}{*}{\textbf{RoG}}         & \textbf{Clean}    & 85.87         & 70.34        & 74.00              & 76.06           & 79.61           & 45.76      
                                                          & 64.94         & 54.86        & 56.64              & 57.57           & 56.87           & 47.18        \\
                                      & \textbf{Rand}     & 84.46         & 67.38        & 70.64              & 74.03           & 77.33           & 40.91                                                           & 56.36         & 49.83        & 51.26              & 51.84           & 51.97           & 42.91        \\
                                      & \textbf{Ours}     & \textbf{75.61}  & \textbf{38.06}$_{\mathbf{\color{red}46\%\downarrow}}$  & \textbf{34.16} & \textbf{63.86} & \textbf{48.46}$_{\mathbf{\color{red}39\%\downarrow}}$ & \textbf{8.85}$_{\mathbf{\color{red}81\%\downarrow}}$                                                    & \textbf{47.95}  & \textbf{36.45}$_{\mathbf{\color{red}34\%\downarrow}}$  & \textbf{35.95} & \textbf{43.58} & \textbf{39.96}$_{\mathbf{\color{red}30\%\downarrow}}$ & \textbf{26.71}$_{\mathbf{\color{red}43\%\downarrow}}$ \\ 
\midrule
\multirow{3}{*}{\textbf{GCR}}         & \textbf{Clean}    & 89.80         & 71.07        & 77.56              & 74.61           & 78.50           & 44.29                                                                                                 & 62.50         & 52.31        & 52.24              & 57.34           & 52.96           & 36.82          \\
                                      & \textbf{Rand}     & 88.70         & 70.11        & 76.86              & 73.07           & 78.01           & 43.98                                                           & 60.04         & 50.02        & 50.01              & 54.82           & 51.18           & 35.15                \\
                                      & \textbf{Ours}     & \textbf{85.75}  & \textbf{63.94}$_{\mathbf{\color{red}10\%\downarrow}}$ & \textbf{69.52} & \textbf{69.63} & \textbf{69.10}$_{\mathbf{\color{red}12\%\downarrow}}$ & \textbf{34.52}$_{\mathbf{\color{red}22\%\downarrow}}$                                                    & \textbf{57.46}  & \textbf{46.93}$_{\mathbf{\color{red}10\%\downarrow}}$ & \textbf{46.73} & \textbf{52.06} & \textbf{47.35}$_{\mathbf{\color{red}11\%\downarrow}}$ & \textbf{31.44}$_{\mathbf{\color{red}15\%\downarrow}}$ \\ 
\midrule
\multirow{3}{*}{\textbf{G-retriever}} & \textbf{Clean}   & 72.36         & 52.39       & 66.42              & 54.47           & 64.47           & 31.88                                                                                                  & --             & --           & --                  & --               & --               & --                    \\
                                      & \textbf{Rand}    & 72.05         & 52.20       & 68.55              & 50.88           & 64.31           & 31.63                                                            & --            & --           & --                  & --               & --               & --                    \\
                                      & \textbf{Ours}    & \textbf{67.38}     & \textbf{46.31}$_{\mathbf{\color{red}12\%\downarrow}}$ & \textbf{54.92} & \textbf{47.96} & \textbf{59.58}$_{\mathbf{\color{red}10\%\downarrow}}$ & \textbf{25.86}$_{\mathbf{\color{red}19\%\downarrow}}$                                                   & -- & -- & -- & -- & -- & -- \\ 
\midrule
\multirow{3}{*}{\textbf{SubgraphRAG}} & \textbf{Clean}   & 80.28         & 66.94       & 73.96              & 68.05           & 76.66           & 51.29                                                                                                  & 56.61         & 49.74       & 52.63              & 50.99           & 53.87           & 41.57                \\
                                      & \textbf{Rand}    & 79.73         & 66.06       & 73.30              & 67.20           & 76.29           & 49.69                                                            & 54.12         & 47.38       & 50.07              & 48.81           & 51.49           & 39.51                \\
                                      & \textbf{Ours}    & \textbf{75.98}    & \textbf{50.22}$_{\mathbf{\color{red}25\%\downarrow}}$ & \textbf{55.08} & \textbf{62.14} & \textbf{66.15}$_{\mathbf{\color{red}14\%\downarrow}}$ & \textbf{27.52}$_{\mathbf{\color{red}46\%\downarrow}}$                                                  & \textbf{50.86}    & \textbf{41.82}$_{\mathbf{\color{red}16\%\downarrow}}$ & \textbf{43.58} & \textbf{45.28} & \textbf{46.25}$_{\mathbf{\color{red}14\%\downarrow}}$ & \textbf{32.91}$_{\mathbf{\color{red}21\%\downarrow}}$ \\ 
\bottomrule
\end{tabular}
}
\label{tab:main1}
\end{table*}

\subsection{Baseline and Implementation Details}
As there is no prior work on data poisoning attacks targeting KG-RAG systems, we compare our proposed method with a structure-preserving \textbf{random modification} baseline. For each question, we identify triples in the KG that involve the question entity and construct corrupted variants by randomly replacing the other entity. 

To generate adversarial answers, we employ OpenAI's GPT-4~\cite{achiam2023gpt} and conduct multiple rounds of generation to produce $N = 5$ candidates for each question.
To instantiate the generation of relation paths, we adopt LLM$_\text{RoG}$, a publicly available\footnote{https://huggingface.co/rmanluo/RoG} KG-specific language model based on LLaMA-2-7B-hf~\cite{touvron2023llama}. As mentioned in Section~\ref{sec:rel_path}, it fits well with our needs. The prompt template used to generate relation paths is shown in Figure~\ref{fig:prompt2}. We set the perturbation budget to $K=4$ per adversarial answer during attacks. Given $N=5$ adversarial answers per question, the total number of injected triples is capped at $N \times K = 20$ per question, which is 
notably fewer than the average number of triples in the original question-associated subgraphs on both WebQSP and CWQ (see Table~\ref{tab:dataset1}). For fair comparison, the \textbf{random modification} baseline is also constrained to insert 20 corrupted triples per question.


We evaluate the impact of the attack across four representative KG-RAG methods described in Section~\ref{sec: kg-rag_methods}: RoG, GCR, G-retriever, and SubgraphRAG. For GCR and SubgraphRAG, which report performance under different LLM backends, we use GPT-3.5-turbo for consistency and fair comparison. For RoG and G-retriever, which rely on fine-tuned internal LLMs, we follow their original implementations. Further analysis of the impact of different LLM choices is also provided in Section~\ref{sec:diff_llm}.

\subsection{Overall Attack Results}
To evaluate the effectiveness of our proposed data poisoning attack on KG-RAG methods, we conduct a comprehensive analysis across KGQA performance before and after attacks. The results are summarized in Table~\ref{tab:main1}, and our analysis focuses on following aspects:

\noindent \textbf{Degradation of KGQA Performance:}  
Compared to random manipulation to poison the KG, which results in marginal performance degradation, our attack induces substantial and consistent declines across all standard QA metrics for KG-RAG methods.
For example, on the WebQSP dataset, the F1 score of RoG drops from 70.34\% to 38.06\%, a relative decrease of 46\%. 
Similar trends are consistently observed for the other KG-RAG methods on both datasets, indicating that these four KG-RAG methods are highly vulnerable to knowledge poisoning: even minor KG corruption can disrupt retrieval quality and lead to noticeable declines in QA performance.

In addition, we observe that the performance decline is particularly noticeable on the Hits@1 and EM metrics. Our attack strategy not only impairs general answer quality but also causes KG-RAG methods to prioritize incorrect over correct answers.
The substantial drop in EM further demonstrates that our attack strategy is highly effective in flipping originally correct predictions. Specifically, RoG’s EM value on WebQSP decreases from 45.76\% to 8.85\%, implying that approximately 81\% of the questions that were originally answered correctly have been corrupted under our attack. Similarly, SubgraphRAG experiences an EM reduction of around 46\%, highlighting that even high-performing KG-RAG methods under clean settings remain vulnerable to data poisoning.

\noindent \textbf{Dataset-Specific Susceptibility:} 
Comparing the KGQA performance of the same KG-RAG method between two benchmarks, we find that these four KG-RAG methods suffer more severe performance degradation on WebQSP than on CWQ. 
For example, considering the metric EM, our attack leads to a remarkable 81\% relative drop for RoG on WebQSP, which is nearly twice the decline observed on CWQ, and SubgraphRAG exhibits a similar trend, with a 65\% drop on WebQSP compared to a 34\% drop on CWQ. 
This difference is partly due to the inherent difficulty of CWQ. 
As indicated by the Hit scores in the Clean setting without attacking, KG-RAG methods are only able to provide at least one correct answer for around 60\% of the questions in CWQ, which is substantially lower than their performance on WebQSP. This reflects the inherent difficulty of CWQ, which in turn limits the extent to which performance can further decrease under attack. 
In contrast, WebQSP contains a larger proportion of questions that can be answered with little difficulty, making it more susceptible to targeted poisoning.

\noindent \textbf{Robustness of KG-RAG Methods:} 
We further investigate the robustness of different KG-RAG methods. 
Among methods based on subgraph retrieval, although G-retriever appears relatively more robust, its low clean performance leaves less room for further degradation.
In contrast, SubgraphRAG exhibits greater vulnerability to attacks but maintains moderately better performance than G-retriever under both clean and attacked scenarios.
For KG-RAG methods that retrieve reasoning paths as evidence, GCR demonstrates much higher robustness than RoG. A possible explanation is that RoG fully relies on a KG-specific LLM for answer generation, whereas GCR uses the KG-specific LLM only to propose candidate answers, and delegates the final answer selection to a general-purpose LLM. 
This two-stage design, which separates candidate generation and answer selection between two LLMs, likely mitigates the impact of poisoned knowledge on GCR's final predictions. 
These findings indicate that the choice and functional role of LLMs within KG-RAG systems critically influence their robustness to data poisoning, a factor we analyze further in Section~\ref{sec:diff_llm}.
Generally, comparing these four KG-RAG methods, RoG suffers server QA performance degradation under attack. SubgraphRAG performs moderately well but is more vulnerable to knowledge poisoning. G-retriever shows apparently competitive robustness, with only 12\% relative drop in F1 score on WebQSP, and GCR achieves a favorable balance, maintaining both strong robustness and high answer quality under attack.

From above observations, we find that \textbf{the four recent and representative KG-RAG methods are highly vulnerable to knowledge poisoning attacks, where the attack effectiveness varies across different benchmarks and methods.} These results highlight the need to develop robust KG-RAG methods, either against knowledge poisoning threats or under low data quality.

\begin{table}[]
\caption{Results ($\%$) of adversarial manipulation on WebQSP and CWQ.}
\fontfamily{ptm}\selectfont
\resizebox{0.48\textwidth}{!}{
\fontsize{12}{16}\selectfont
\begin{tabular}{cc|ccc}
\toprule
\textbf{KG-RAG} & \textbf{Dataset}  & \textbf{A-Precision} & \textbf{A-H@1} & \textbf{A-MRR} \\ 
\midrule
\multirow{2}{*}{\textbf{RoG}} & \textbf{WebQSP}  & 49.66 & 42.75 & 56.34 \\ 
& \textbf{CWQ}  & 46.10 & 43.19 & 48.12 \\ 
\midrule
\multirow{2}{*}{\textbf{GCR}} & \textbf{WebQSP}  & 13.52 & 14.43 & 18.11 \\
& \textbf{CWQ}  & 18.41 & 19.09 & 21.51 \\ 
\midrule
\multirow{2}{*}{\textbf{SubgraphRAG}} & \textbf{WebQSP}  & 27.02 & 18.86 & 26.61 \\
& \textbf{CWQ} & 30.30 & 29.00 & 31.11 \\ 
\midrule
\multirow{2}{*}{\textbf{G-retriever}} & \textbf{WebQSP} & 12.96 & 15.01 & 16.58 \\
& \textbf{CWQ} & - & - & - \\ 
\bottomrule
\end{tabular}}
\label{tab:attack-results}
\end{table}

\begin{figure}[h]
  \centering
  \includegraphics[width=0.98\linewidth]{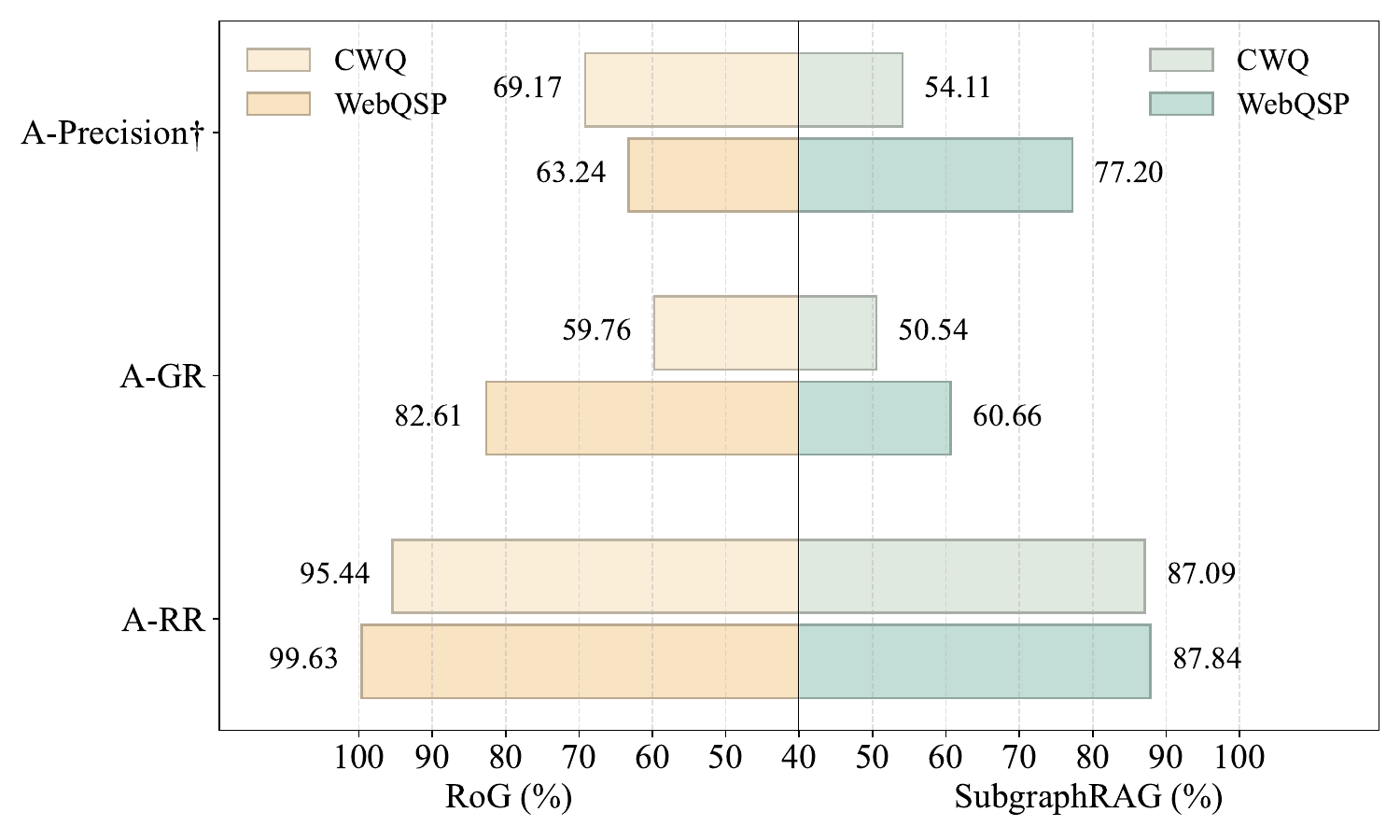}
  \caption{Attack impact on different stages of KG-RAG systems.}
  \label{fig:but}
\end{figure}

\subsection{Effectiveness of Adversarial Manipulation}
To further analyze the attack effectiveness beyond QA degradation, we examine how the assigned adversarial answers are ultimately generated by KG-RAG methods. 

As shown in Table~\ref{tab:attack-results}, RoG is easily misled to produce targeted negative answers. Moreover, the high A-MRR values (56.34\% and 48.12\% on two benchmarks) of RoG suggest its limited ability to distinguish adversarial answers from plausible ones. In contrast, GCR and SubgraphRAG exhibit stronger resistance to the misleading towards adversarial answers.
As for G-retriever, the lower attack efficiency may stem from its generation mechanism, where the retrieved subgraph is encoded into embeddings and concatenated as a prefix to the input text embeddings of the LLM, and then the final decoded predictions are obtained. This probably makes the attack less controllable, compared to directly presenting adversarial content as natural language input via in-context learning. Interestingly, although GCR and SubgraphRAG show smaller QA performance degradation on CWQ than on WebQSP (shown in Table~\ref{tab:main1}), they actually present higher values across A-Precision, A-H@1, and A-MRR on CWQ. 

These results are consistent with the observations in Table~\ref{tab:main1}, confirming that the QA degradation primarily results from our proposed attack, which effectively induces the generation of adversarial answers, rather than being attributed to the hallucinations of LLMs themselves.

\begin{figure*}[t]
    \centering
    \subfloat[KGQA performances of RoG]{
        \includegraphics[width=0.235\textwidth]{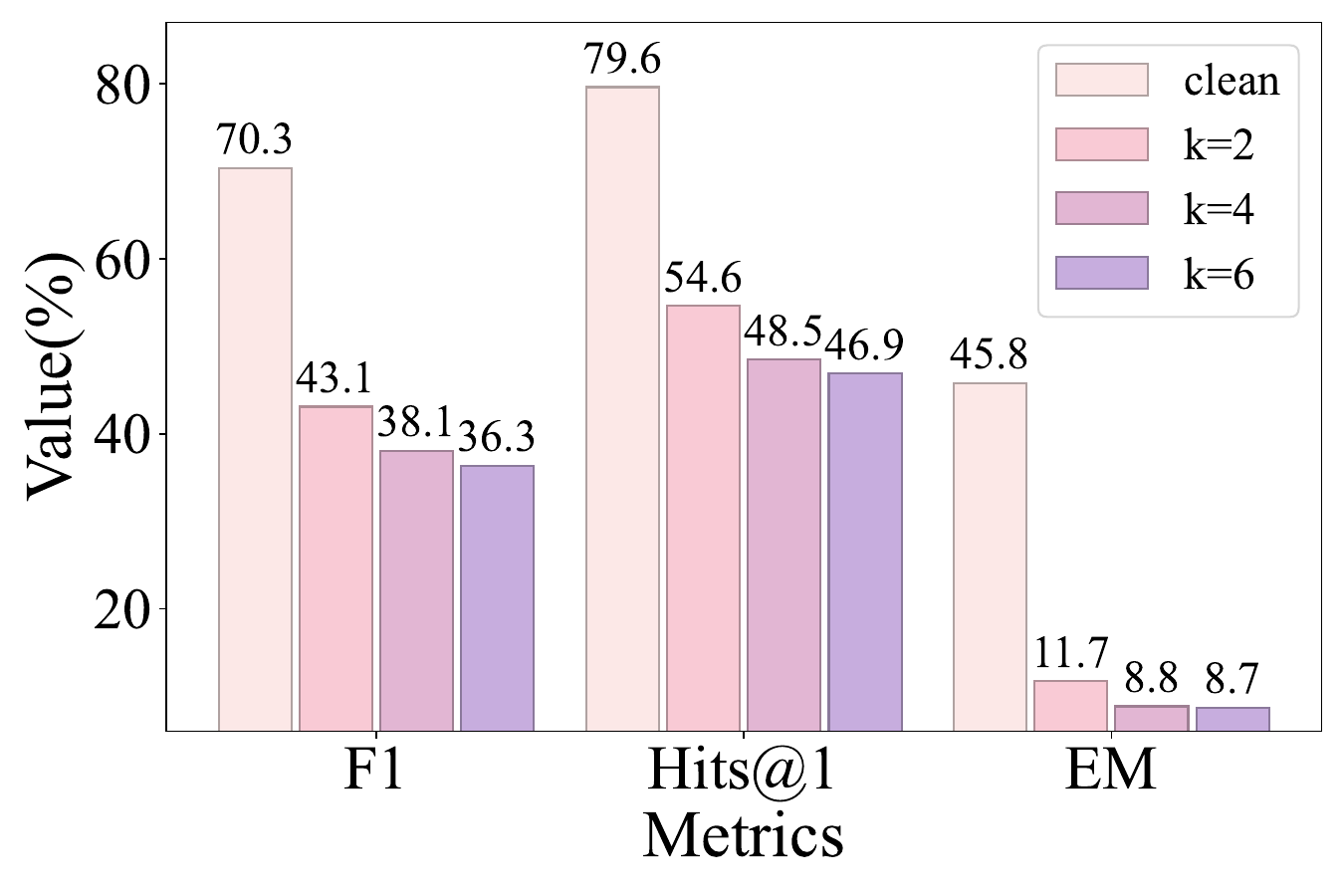}
    }
    \hfill
    \subfloat[Adversarial Results of RoG]{
        \includegraphics[width=0.235\textwidth]{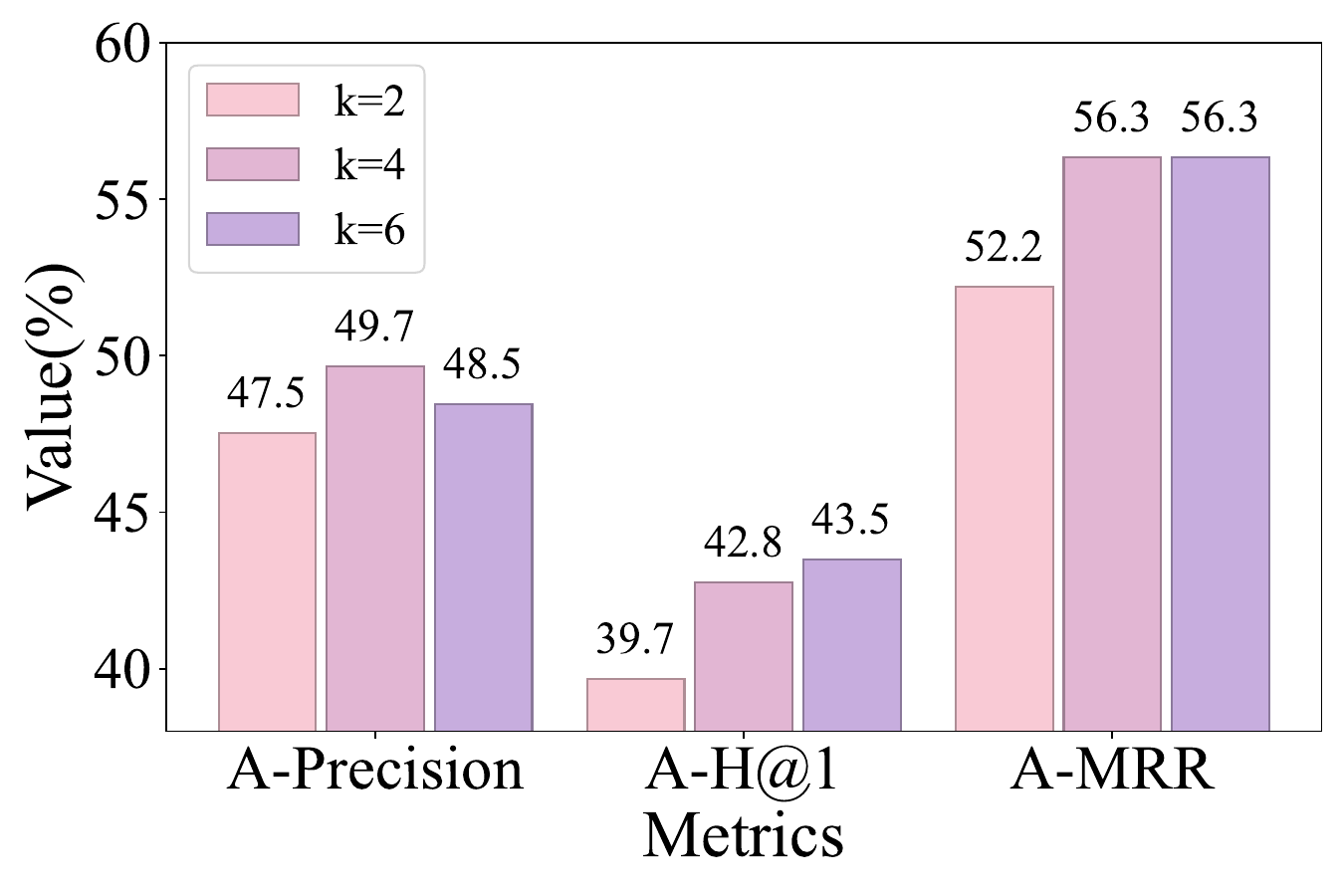}
    }
    \hfill
    \subfloat[KGQA performances of GCR]{
        \includegraphics[width=0.235\textwidth]{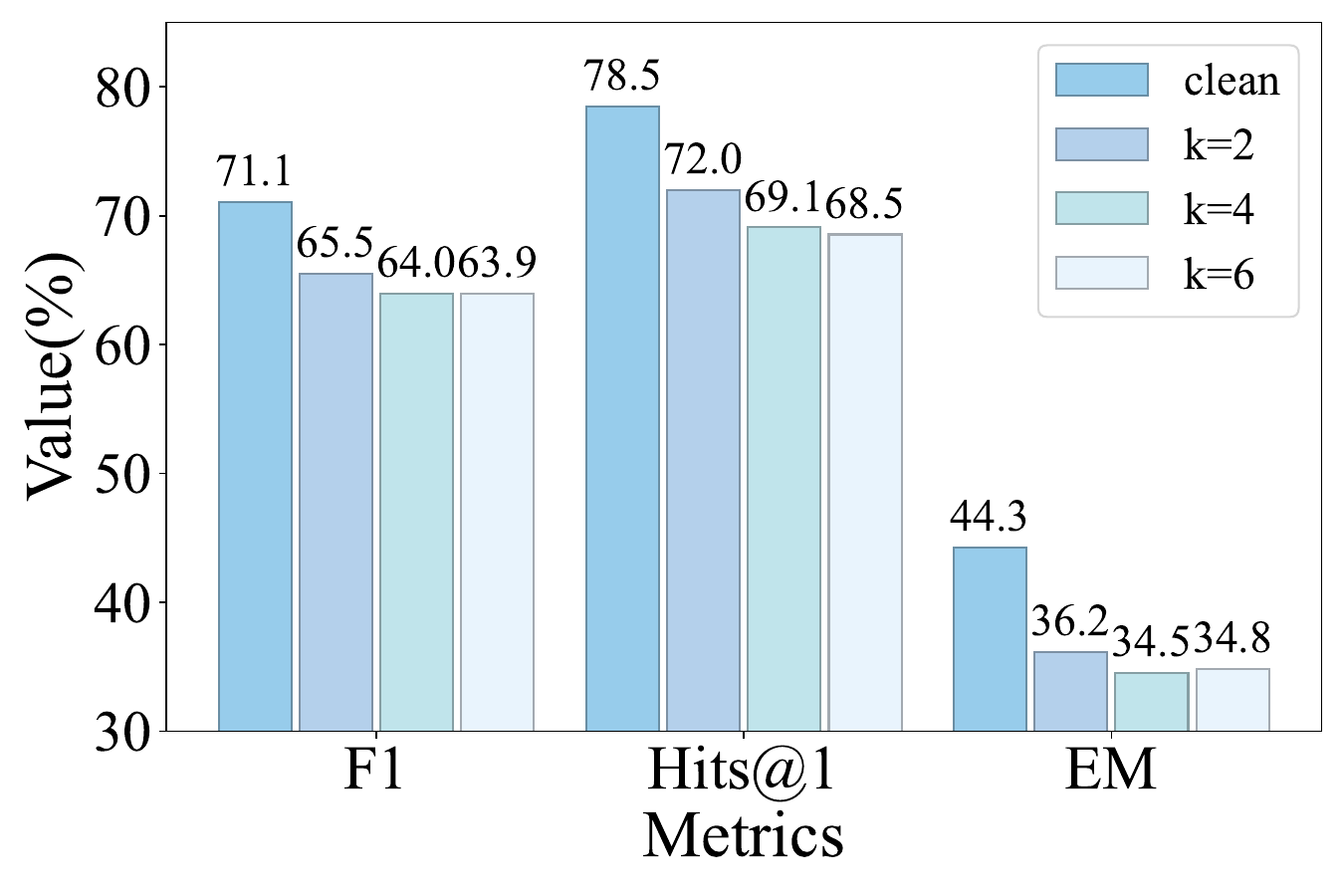}
    }
    \hfill
    \subfloat[Adversarial Results of GCR]{
        \includegraphics[width=0.235\textwidth]{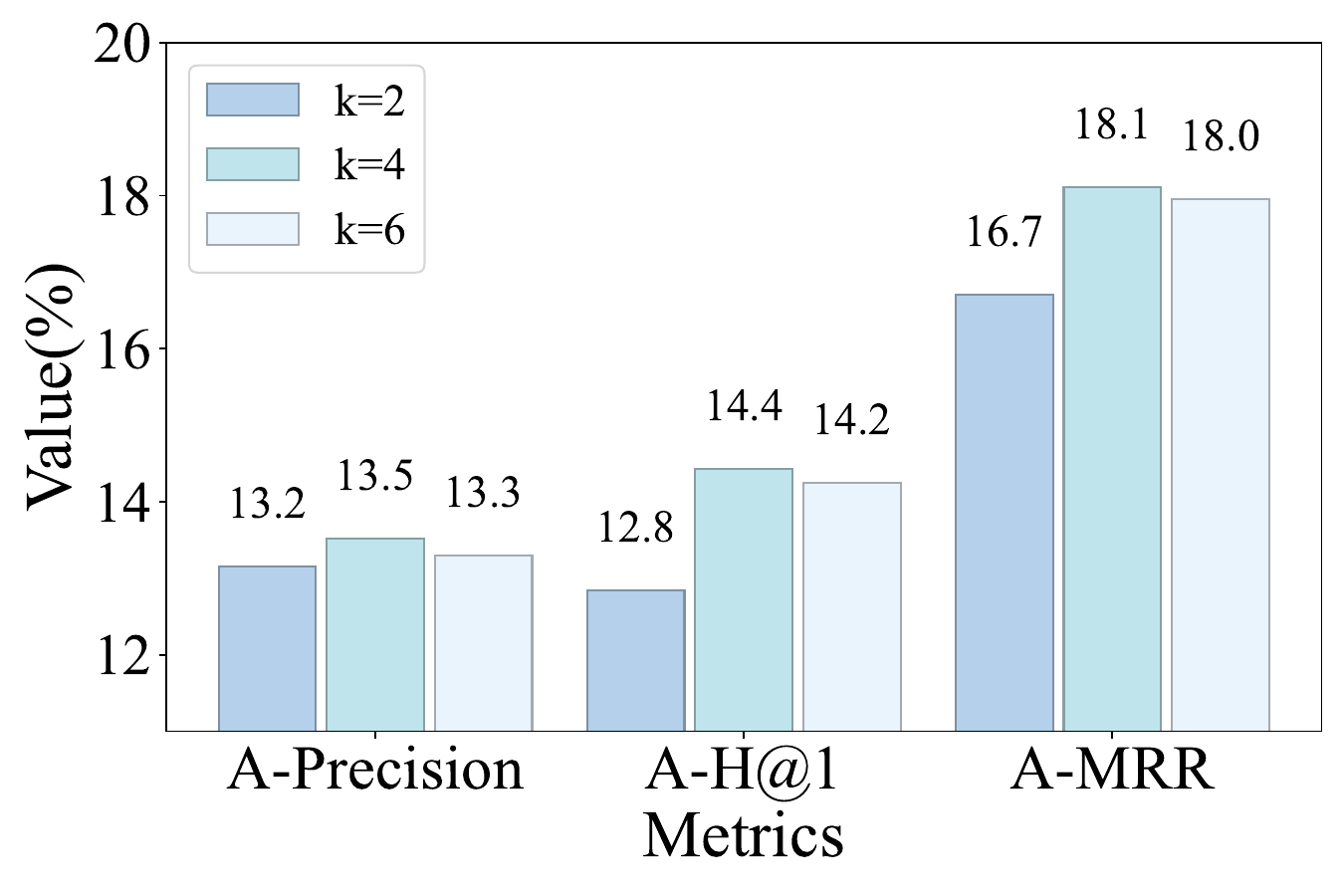}
    }
    \caption{Attack Effectiveness across increasing poisoning scales.}
    \label{fig:perturb_triple_num}
\end{figure*}

\begin{table*}[ht]
\caption{Performance of RoG using different LLMs, with and without data poisoning attack.}
\fontfamily{ptm}\selectfont
\resizebox{0.98\textwidth}{!}{%
\fontsize{11}{16}\selectfont
\begin{tabular}{llccccccccccc}
\toprule
      &    & \multicolumn{3}{c}{\textbf{F1}}               &  & \multicolumn{3}{c}{\textbf{Hits@1}}                   &  & \multicolumn{3}{c}{\textbf{EM}}                   \\ 
\cmidrule{3-5} \cmidrule{7-9} \cmidrule{11-13} 
\multirow{-2}{*}{\textbf{Category}}      & \multirow{-2}{*}{\textbf{LLM for Generation}}    & \textbf{Clean} & \textbf{Attack}    & \textbf{Degradation}    & \textbf{}                     & \textbf{Clean}            & \textbf{Attack}     & \textbf{Degradation} & \textbf{}                     & \textbf{Clean}            & \textbf{Attack} & \textbf{Degradation} \\ 
\midrule
\multirow{2}{*}{\textbf{Open-source}}     &      \textbf{Llama-2-7B-chat-hf}                & 41.00   & 22.78   & 18.22 (44.43$\downarrow$)   &   & 46.56     & 26.78   & 19.78 (44.48$\downarrow$)    &   & 7.49   & 0.25     & 7.24 (96.66$\downarrow$)             \\
                                          &      \textbf{Llama-3.1-8B-Instruct}          & 54.20   & 32.02   & 22.18 (40.92$\downarrow$)   &   & 32.49     & 27.70   & 4.79 (14.72$\downarrow$)    &   & 26.78  & 5.77     & 21.01 (78.45$\downarrow$)         \\
\midrule
\multirow{2}{*}{\textbf{Closed-source}}   & \textbf{GPT-3.5-turbo}                & 56.51   & 36.19    & 20.34 (35.95$\downarrow$)   &   & 64.13      & 35.75  & 28.38 (44.25$\downarrow$)   &  & 31.02   & 3.87     & 27.15 (87.52$\downarrow$)            \\
                                          & \textbf{GPT-4}                        & 61.33   & 37.31    & 24.02 (39.17$\downarrow$)   &   & 61.36      & 32.49  & 28.87 (47.05$\downarrow$)   &  & 41.34   & 8.91     & 32.43 (78.45$\downarrow$)             \\
\midrule
\textbf{Mixture‑of‑Experts}               & \textbf{DeepSeek-V3-0324}            & 57.69    & 33.00    & 24.69 (42.80$\downarrow$)   &   & 67.04     & 37.70  &  29.34 (43.76$\downarrow$)   &  & 41.94  & 11.81     & 30.13 (71.84$\downarrow$)           \\
\midrule
\textbf{KG-specific}                      & \textbf{LLM$_\text{RoG}$}                        & 70.34     & 38.06    & 32.28 (45.89$\downarrow$)    &   & 79.61     & 48.46   &   31.15 (39.13$\downarrow$)  &  & 45.76   & 8.85    &  36.91 (80.66$\downarrow$)             \\ 
\bottomrule
\end{tabular}%
}
\label{tab:llm-robustness}
\end{table*}

\subsection{Attack Analysis on KG-RAG Stages}
To better understand how our data poisoning attack affects different stages of the KG-RAG methods, we conduct a detailed analysis focusing on how adversarial answers are retrieved and ultimately involved in the final responses.

Specifically, we calculate the proportion of questions in which the retrieved result contains the injected perturbation triples that can lead to adversarial answers,
and denote it as Attack Retrieved Ratio (A-RR). 
To quantify the impact on the generation stage, we focus on the questions of which adversarial answers have been successfully retrieved, and among them, measure the proportion of those questions in which adversarial answers are finally generated, denoted as Attack Generated Ratio (A-GR).  
Besides, we further calculate a more focused measurement of A-Precision, denoted as A-Precision\dag, calculating the A-Precision values over the questions where at least one adversarial answer is generated.

We consider two representative KG-RAG methods, RoG and SubgraphRAG, which retrieve reasoning paths and subgraphs, respectively.  
As illustrated in Figure~\ref{fig:but}, our data poisoning attack achieves high coverage at the retrieval stage, with both methods exhibiting high A-RR scores across both datasets.  
Compared to the success in polluting retrieval results, the KG-RAG methods behavior more resiliently in answer generation.
For example, only about half of the CWQ questions with retrieved adversarial answers are ultimately adopted by SubgraphRAG in the final response. However, for those questions in which adversarial answers are successfully introduced into the final prediction, such answers typically occupy a large proportion of the prediction across both RoG and SubgraphRAG on the two benchmarks.

Based on these results, we observe that \textbf{the retrieval stage constitutes a critical vulnerability in existing KG-RAG methods, underscoring the importance of reliable evidence to secure KG-RAG methods. While the generation stage with LLMs is more robust against poisoned evidence, it still suffers from a polarized response: adversarial content is either fully rejected or dominates the final prediction once adopted.}

\subsection{Attack Analysis on Poisoning Scales}
In this section, we investigate the impact of the poisoning scale by varying the number of poisoned triples $k$ associated with each targeted adversarial answer. 
As shown in Figure~\ref{fig:perturb_triple_num}, larger values of $k$ generally lead to stronger attack effects on both QA accuracy and adversarial manipulation. This is primarily reflected in the improved ability to promote adversarial answers and elevate their ranking, as indicated by higher MAPR and AH@1 scores.

However, the effects of further increasing $k$ beyond $k=4$ diminish, with both QA and attack metrics remaining largely stable. For example, when $k=4$, RoG already fails to produce most of the exactly correct answers, leaving little room for further degradation. In such cases, additional poisoned triples primarily increase the likelihood that the targeted negative answers appear at higher ranks, but can hardly lead to further drops in the overall QA performance of KG-RAG methods. Instead, the increasing poisoning scales would potentially raise the risk of attack detection.

These findings suggest that \textbf{ KG-RAG methods are highly sensitive to knowledge poisoning, as even small-scale perturbations can cause notable performance degradation. Compared to the marginal enhancement of attack effectiveness, increasing poisoning scales may pose a greater concern regarding attack stealthiness, especially considering that large-scale perturbations are often impractical in real-world scenarios.
}

\begin{table*}[t]
\fontfamily{ptm}\selectfont
\caption{Effect of data poisoning on retrieval and prediction: a real example of RoG.}
\centering
\begin{tabularx}{\textwidth}{
  >{\centering\arraybackslash}m{1.8cm} |
  >{\raggedright\arraybackslash}X 
}
\toprule
Question & What language is spoken in the location that appointed Michelle Bachelet to a governmental position speak? \\
\midrule
Ground Truth & Aymara language, Mapudungun Language, Rapa Nui Language, {\color{blue}Spanish Language}, Puquina Language   \\
\midrule
Before Attack &  \textbf{Retrieval:} \newline
1. Michelle Bachelet $\rightarrow$ people.person.nationality $\rightarrow$ Chile $\rightarrow$ \mbox{language.human\_language.countries\_spoken\_in} $\rightarrow$ Spanish Language \newline
\textbf{Prediction:} \newline
{\color{blue}Spanish Language}   \\
\midrule
Adversarial Targets & Portuguese, Italian, French, German, Dutch \\
\midrule
After Attack  &  \textbf{Retrieval:} \newline
1. Michelle Bachelet $\rightarrow$ people.person.nationality $\rightarrow$ Germany $\rightarrow$ \mbox{location.country.languages\_spoken} $\rightarrow$ {\color{red}German} \newline
2. Michelle Bachelet $\rightarrow$ people.person.nationality $\rightarrow$ Chile $\rightarrow$ \mbox{location.country.languages\_spoken} $\rightarrow$ {\color{red}Italian} \newline
3. Michelle Bachelet $\rightarrow$ people.person.nationality $\rightarrow$ Chile $\rightarrow$ \mbox{location.country.languages\_spoken} $\rightarrow$ Spanish Language \newline
4. Michelle Bachelet $\rightarrow$ people.person.nationality $\rightarrow$ Chile $\rightarrow$ \mbox{location.country.official\_language} $\rightarrow$ {\color{red}Portuguese} \newline
5. Michelle Bachelet $\rightarrow$ people.person.nationality $\rightarrow$ Chile $\rightarrow$ \mbox{location.country.official\_language} $\rightarrow$ {\color{red}French} \newline
6. Michelle Bachelet $\rightarrow$ people.person.nationality $\rightarrow$ Chile $\rightarrow$ \mbox{location.country.official\_language} $\rightarrow$ {\color{red}German} \newline
7. Michelle Bachelet $\rightarrow$ people.person.nationality $\rightarrow$ Chile $\rightarrow$ \mbox{location.country.official\_language} $\rightarrow$ {\color{red}Dutch}  \newline 
\textbf{Prediction:} \newline
{\color{red}Italian Language, German Language, French,} {\color{blue}Spanish Language}  \\
\bottomrule
\end{tabularx}
\label{tab:case_study}
\end{table*}

\subsection{Attack Analysis on KG-RAG Generators}
\label{sec:diff_llm}
In this section, we investigate the robustness of different LLMs used in the generation stage of KG-RAG methods under our data poisoning attacks. We focus on the KG-RAG method of RoG and adopt different LLMs used for final answer generation. Specifically, we evaluate four categories of LLMs: open-source models (LLaMA2-7B-chat-hf, LLaMA3.1-8B-Instruct), closed-source models (GPT-3.5-turbo, GPT-4), mixture-of-experts (MoE) models (DeepSeek-V3-0324), and KG-specific models (the RoG-integrated LLM). 
The retrieval results and prompt templates are kept consistent across all settings, with minimal formatting adjustments for each LLM to accommodate their input formatting requirements. 

Based on the results in Table~\ref{tab:llm-robustness}, we have following observations: 
(1). The KG-specific LLM (LLM$_\text{RoG}$) achieves the greatest QA performance without data poisoning attack, but also suffers the most significant degradation under attack.  For example, its F1 score drops by 45.89\% relative to its clean performance, a decline greater than that of any other LLM in comparison.
(2). Closed-source LLMs (GPT-3.5 and GPT-4) and the MoE model (DeepSeek-V3-0324) consistently outperform open-source LLaMA variants both before and after attack. Notably, LLaMA-2-7B-chat-hf exhibits the weakest robustness, with its EM score nearly eliminated under attack. Although LLaMA-3.1-8B-Instruct shows improved robustness, its performance remains far below that of GPT-4 and DeepSeek-V3-0324.
(3). With attack, GPT-4 and DeepSeek-V3-0324 demonstrate competitive QA performance to the KG-specific model across several metrics. In particular, DeepSeek-V3-0324 achieves the highest EM score after attack (11.81), surpassing LLM$\text{RoG}$ (8.85). In terms of relative degradation, these models seem to exhibit better robustness than the KG-specific LLM. However, the robustness gaps remain marginal.

\subsection{Case Study}
We present a representative case to provide an explicit illustration of how data poisoning affects both retrieval and generation in KG-RAG systems. As shown in Table~\ref{tab:case_study}, our attack successfully leads to adversarial paths retrieved by RoG, which in turn results in incorrect predictions aligned with the attack targets. 
The attack also leads to \textit{Spanish Language}, the only originally correct answer, being no longer prioritized under attack during answer generation.
Besides, we also observe that several adversarial targets, like \textit{Dutch} and \textit{Portuguese}, although successfully retrieved, are ultimately excluded from the final prediction. This reveals that \textbf{ LLMs in RAG systems do not fully rely on all retrieved content and exhibit a potential implicit preference in content selection, highlighting the need to better understand the underlying such selection behavior in future work.}

\section{Related Works}
\subsection{Attack to RAG}
With the increasing adoption of retrieval-augmented generation (RAG) systems, concerns over their safety have been growing~\cite{ni2025towards}, and adversarial attacks have emerged as a promising approach to study robustness and vulnerabilities of RAG systems.
Recent efforts to attack retrieval-augmented generation (RAG) systems mainly focus on manipulating the retrieval corpus to induce malicious responses.
PoisonedRAG \cite{zou2024poisonedrag} conducts the first investigation to inject crafted passages that are retrievable for specific questions and steer the LLM toward attacker-desired answers. While effective, it is limited to fixed question-answer pairs. BadRAG \cite{xue2024badrag} generalizes this approach to semantically grouped questions by using contrastive optimization, enabling broader yet still topic-bound attacks. It also explores indirect generative threats such as denial-of-service and sentiment bias. A recent study~\cite{tan-etal-2024-glue} proposes a composite adversarial document structure that includes three components: an Adversarial Target Sequence, an Adversarial Retriever Sequence, and an Adversarial Generation Sequence. This design enables joint attacks on both the retriever and generator components of RAG systems, optimizing adversarial effectiveness through a dual-objective strategy. 

Although these studies demonstrate that RAG systems are vulnerable to adversarial manipulation of the external knowledge source, they primarily focus on traditional RAG systems that retrieve unstructured texts. However, with the growing incorporation of structured knowledge sources, typically KGs, the security implications of such systems remain largely unexplored.

\subsection{KG Poisoning}
Poisoning attacks on knowledge graphs (KGs) have also received increasing attention, although most prior work focuses on their impact on knowledge graph embedding (KGE) models.
These studies~\cite{ijcai2019p674, banerjee2021stealthy, you2023mass} typically conduct targeted attacks by injecting adversarial triples into the KG structure, aiming to degrade link prediction performance for specific target entities or relations. In contrast to these targeted approaches, recent work~\cite{10.1145/3626772.3657702} explores untargeted KG poisoning by leveraging logic rules to identify and manipulate high-impact triples that significantly affect the global semantics of the KG. These studies, despite focusing on KGE performance, demonstrate that the semantics of KGs are fragile to be polluted, which may further affect KG-based applications such as recommendation and KG-RAG systems. Thus, investigating the influence of KG poisoning on KG-RAG systems is both timely and urgent for understanding their security implications, as KG-RAG has emerged as a popular extension of RAG systems.

\section{Conclusion}
In this work, we present the first study on the safety risks of KG-RAG, a popular extension of the RAG framework. We explore knowledge poisoning attacks and propose a practical scenario in which the attacker is restricted to injecting perturbations to pollute KGs, with the goal of misleading the reasoning results of KG-RAG systems. We develop an attack strategy that inserts crafted triples that can complete adversarial reasoning paths, which may mislead the retriever or generator into producing incorrect answers. Experimental results on two benchmarks and four representative KG-RAG models demonstrate the effectiveness of the proposed attack.

Beyond overall attack success, our analysis reveals the following key insights:

\noindent (1). The safety of KG-RAG systems is particularly susceptible to knowledge poisoning attacks, and their structural vulnerability primarily resides in the retrieval stage. Retrievers in existing methods lack effective mechanisms to distinguish or exclude poisoned triples, allowing adversarial information to flow into the generation stage. Consequently, the overall robustness of KG-RAG systems often relies heavily on the LLMs' persistence to noisy content.

\noindent (2). Different LLMs exhibit varying levels of resistance to noise. While equipping LLMs with structured KG knowledge can improve their performance on KG-related tasks,  it also tends to make them more vulnerable to adversarial KG manipulation. An improving direction may lie in combining the strengths of both KG-specific and general-purpose LLMs, aiming to maintain high answer accuracy while enhancing resistance to poisoned knowledge.

\noindent (3). Increasing the number of injected triples does not guarantee a consistent gain in attack effectiveness. Meanwhile, LLMs employed in KG-RAG generate stages may not fully utilize all retrieved content and could exhibit implicit preferences during answer generation, which needs further explora.

As future work, we aim to enhance the attack effectiveness by inserting perturbations that are more consistent with the LLM's knowledge, thereby increasing the likelihood that retrieved adversarial triples are incorporated into the final answer. We also plan to investigate deletion and modification attacks to develop a broader understanding of RAG safety.










\bibliographystyle{elsarticle-num}

\bibliography{cas-refs}



\end{document}